\documentclass[12pt]{article}
\usepackage[bottom=2cm, top=2cm, left=2cm, right=2cm]{geometry}
\usepackage{color}

\usepackage{subfigure}
\usepackage{amsmath}
\usepackage[utf8]{inputenc}
\usepackage{standalone}
\usepackage{tikz}
\usepackage{amssymb}
\usepackage{eurosym}
\usepackage{hyperref}
\usepackage{caption}
\definecolor{b}{rgb}{0,0,.4}	
\definecolor{g}{rgb}{0,.3,0}	
\definecolor{n}{rgb}{0,0,0}	
\definecolor{h}{rgb}{0.4,0.2,0.2}	
\definecolor{v}{rgb}{0.2,0.6,0}

\newcommand{\C}{{\mathbb C}}

\newcommand{\E}{{\mathbb E}}

\newcommand{\N}{{\mathbb N}}

\renewcommand{\P}{{\mathbb P}}

\newcommand{\R}{{\mathbb R}}

\newcommand{\X}{{\mathbb X}}

\newcommand{\II}{{\mathcal{I}}}

\newcommand{\PP}{{\mathcal{P}}}

\newcommand{\WW}{{\mathcal{W}}}

\newcommand{\bX}{\boldsymbol X}
\newcommand{\bY}{\boldsymbol Y}

\newcommand{\bbeta}{\boldsymbol \beta}

\newcommand{\beps}{\boldsymbol \varepsilon}

\newcommand{\bmu}{\boldsymbol \mu}

\DeclareMathOperator*{\argmin}{arg\,min}	

\newcommand{\ov}\overline
\newcommand{\what}{\widehat}
\newcommand{\wtilde}{\widetilde}

\definecolor{gray}{rgb}{0.5,0.5,0.5}
\definecolor{red}{rgb}{0.8,0,0}
\definecolor{dred}{rgb}{0.5,0,0}
\definecolor{blue}{rgb}{0,0.1,1}
\definecolor{dblue}{rgb}{0,0.1,0.6}
\definecolor{cyan}{rgb}{0,0.7,.2}
\definecolor{dcyan}{rgb}{0,0.5,.5}

\usetikzlibrary{arrows,positioning,shadows, calc, intersections, fit, automata}
\usetikzlibrary{arrows.meta}
\usetikzlibrary{decorations.pathreplacing}

 \usepackage{bbm}
\begin{document}

	\begin{center}
	{\large \textbf{X-model: further development and possible modifications}}
	
	Sergei Kulakov\footnote{Corresponding author. University of Duisburg-Essen.  Berliner Platz 6-8, WST-C.11.19, 45127, Essen, Germany. Email: sergei.kulakov@uni-due.de.} 

\end{center}
\begin{abstract}
	Despite its critical importance, the famous X-model elaborated by \cite{ziel2016electricity} has neither bin been widely studied nor further developed. And yet, the possibilities to improve the model are as numerous as the fields it can be applied to. The present paper takes advantage of a technique proposed by \cite{coulon2014hourly} to enhance the X-model. Instead of using the wholesale supply and demand curves as inputs for the model, we rely on the transformed versions of these curves with a perfectly inelastic demand. As a result, computational requirements of our X-model reduce and its forecasting power increases substantially. Moreover, our X-model becomes more robust towards outliers present in the initial auction curves data. 
	
	
\end{abstract}

\section{Introduction}
As has already been accentuated multiple times (for example, see  \cite{boyle2004renewable},  \cite{weron2007modeling} or \cite{spiecker2014future}), the shift to cleaner power is accelerating at a growing pace. Due to their apparent advantages, renewable resources are becoming increasingly competitive. They also are exerting a profound influence upon contemporary energy systems. Intermittent in nature and weather-dependent, green energies have elevated the importance of forecasting over the past several years. More accurate predictions, as is now obvious,  lead to hefty cost reductions and allow the stability of the whole power system to be maintained. 

Immense complexity of energy markets provided forecasters with a vast variety of datasets and variables to study. Therefore, a numerous amount of different forecasting models has emerged over time (see e.g. \cite{bunn2000forecasting} or \cite{weron2014electricity}). One of those models is the so-called X-model developed by \cite{ziel2016electricity}. Despite following a truly unconventional approach, this model has proven to be a very powerful tool for conducting price and volume forecasts in energy markets. The X-model will thus be given a thorough scrutiny in the present study. 

In fact, the core of the X-model is constructed around a relatively simple idea. When attempting to make a price or volume prediction in an electricity market, scientists typically consider price and volume time series. Instead, however, scientists may focus on making a forecast for the entire wholesale supply and demand curves. Since these curves are used to settle equilibrium prices, the intersection of the prognosticated curves will constitute the price or volume forecast. 

To explain the functioning of the model, let us first consider the demand curve. To obtain a prediction for the entire demand curve at time period $t+1$, we first need to select several points on this curve. These points correspond to certain prices and in the original paper by \cite{ziel2016electricity} were referred to as the price classes. Then, we construct a time series-based model for volume forecasting for each of the selected price classes. Combining the obtained forecasts together (or, loosely speaking, drawing "a line" over the predicted points) thus yields a prediction for the entire demand curve at time period $t+1$. Then, we progress similarly to obtain a forecast for the entire supply curve. Afterwards, as has already been mentioned in the previous paragraph, we simply search for the intersection between the two predicted auction curves and conclude that this intersection coincides with our equilibrium price or volume forecast.

Hence, as its major intuition may suggest, the X-model can be used particularly well when it comes to forecasting price spikes. Since the model incorporates best properties of both time series and structural analyses, the X-model is capable of capturing the bidding behavior of market participants more precisely. This ability, in turn, results in more accurate forecast of extreme price events.

Nevertheless, despite being a model of exceptional importance, it has not been widely developed further. This paper aims to fill this gap and proposes presumably the first improvement of the original X-model. The kernel of the present study is based upon the paper written by \cite{coulon2014hourly} who showed that the initial wholesale auction curves can be transformed into their analogues with a perfectly inelastic demand curve. We will show that forecasting accuracy of the X-model improves and its computational burden lessens if the curves are transformed prior to making a prognostication. 

Fundamentally, the inelastic demand curve is the main reason behind a superior performance of the modified X-model. First, predicting only one point instead of the entire demand curve requires a substantially smaller amount of time. Therefore, the modified X-model delivers final results faster than the original X-model. Second, the modified X-model is much less dependent on the outliers present in the original auction curves data. Speaking generally, the initial wholesale demand curve has a very sophisticated composition and shape. Predicting this curve correctly is thus a task of great complexity. Forecasting the inelastic demand is, in turn, much simpler. As a result, the modified X-model is not only quicker, but also more accurate. 

This paper is organized as follows. The next section comments briefly on the used data set and the corresponding manipulations with the data. Section 3 elaborates on the transformation of the auction curves and explains why this transformation leads to a significant improvement of the X-model. Section 4 provides the fundamentals of the underlying time series process. Section 5 is a discussion of the obtained results. Section 6 concludes the paper. 

\section{Data set}
The present study was conducted on the auction curves data from the German EPEX SPOT SE. Additional data sets were obtained from the ENTSOE. These data sets comprise wind and solar power forecasts and the total generation forecast. The in-sample period used in the current paper is from 2016-01-01 to 2017-01-01. The out-of-sample period is the year 2017. Following the regulation of the EPEX, the maximal bid price in the market is equal to $P_{\max}=3000$, the lowest bidding price amounts to $P_{\min}=-500$. The data was clock-change adjusted. The missing hours were calculated using the two values before and after them, whereas the average value of two double hours in October was taken to solve the problem. 

There were two big clusters of outliers present in the original data. The first cluster was detected at the price $P_{\min}$ in the supply curve, the second one at the price $P_{\max}$ in the demand curve. The fact that these clusters are outliers becomes apparent given that observations at points distant from $P_{\min}$ and $P_{\max}$ did not exhibit any peculiarities. Moreover, from an economic standpoint, it is possible that market participants tried to bid unrealistic volumes at the very extremes of the auction curves in a hope to get very profitable deals. 

Please note that these outliers do impede the functioning of the model because they affect the compositions of the forecasted auction curves. Speaking technically, however, we could have taken points e.g. $P_S^1=-495$ and $P_D^1=2995$ to construct the X-model. If we would have done so, the outliers would barely influence the model since prices are almost never realized at the extremes of the auctions curves. From this perspective, we would suffer only a marginal loss in informational efficiency if we would have taken points $P_S^1=-495$ instead of $P_{\min}$ and $P_D^1=2995$ instead of $P_{\max}$. However, we do not want to tolerate this loss and stick to the officially established price bounds. Therefore, the outliers are to be processed.  

To clean these outliers, a suggestion from \cite{weron2007modeling} was taken and a typical expert-type regression model was constructed. The model is similar to \cite{weron2008forecasting} or \cite{ziel2016forecasting} with lags 1,2, and 7 and Monday, Saturday and Sunday. Moreover, the corresponding value at the point $P_S^1=P_{\min}+5$ and $P_D^1=P_{\max}-5$ for the supply and demand curves, respectively, was used as an additional regressor in the model. Since the values at points $P_S^1$ and $P_D^1$ of the same curves were taken into account, the method, though simple, yields credible and precise results. 

\section{Transformation of the auction curves}

As was mentioned earlier, it is possible to transform the actually observed auction curves into their analogues with a perfectly inelastic demand curve. The basics of this transformation are provided in \cite{coulon2014hourly}, whereas possible applications of this theory can be found in \cite{kulakov2019impact} and \cite{kulakov2019determining}. A graphical representation of the transformation can be found in the Figure \ref{FIG5} below. 

\begin{figure}[h]
	\centering
	\vspace{-1cm}
	\scalebox{0.9}{\input{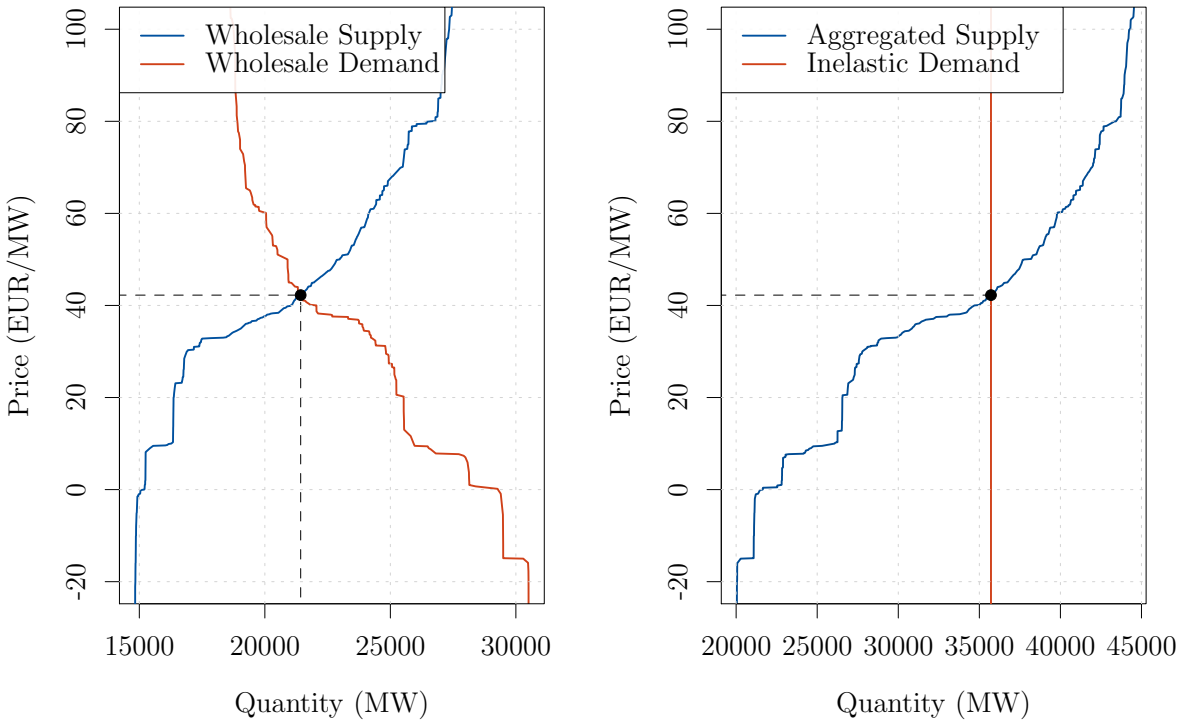}}		
	\caption{{A wholesale market equilibrium in the EPEX SPOT SE on 2017-02-01 at 00:00:00 (left plot) vs. its manipulated form with an inelastic demand curve (right plot)}}
	\label{FIG5}
\end{figure}

It is of crucial importance to note that the price remains the same after the curves have been transformed. The volume sizes, however, increase. Using the X-model for price forecasting on the transformed curves is thus possible without any further modifications or additional comments. However, our volume forecast will not correspond to the wholesale market volumes.
 
Please note that there are three major benefits from transforming the curves prior to applying the X-model. First, recall that the demand curve after the transformation is represented by only one point. Therefore, it is no longer necessary to predict the whole demand curve, but rather only to make a forecast for this single point. This allows a substantial amount of computational time to be spared. Second, the model becomes more robust towards outliers present in the auction curves data. Being more robust towards the outliers, the accuracy of the model increases greatly. This feature of the modified X-model will be expanded in more details in what follows. Third, the spared computational time can be used to select a greater amount of price classes and thus forecast the supply curve more precisely. Hence, as the above description suggests, transforming the auction curves before using the X-model will not only deliver the results speedier, but will also yield results of higher quality.

\section{Model description} 
\subsection{Transformation of the auction curves}
The first step in describing the model is to comment on the way the auction curves can be transformed. As has been mentioned earlier, the formulas were taken from \cite{coulon2014hourly}. Please note that we consider both auction curves as functions of the price. We can thus simplify the notation significantly. Please note that the expression for the inelastic demand curve can be represented as 
\begin{align}
Dem^{inelastic}_t = WS Dem_t^{-1}(P_{\max})
\end{align}
where $WSDem$ denotes the  demand curve in a wholesale market and $P_{\max}$ shows the maximal price at which market participants can bid under the regulation of a power exchange.  The equation for the inverse supply curve reads
{\footnotesize\begin{align}
	Sup_t^{-1}(z) = WSSup_t^{-1}(z) + WSDem^{-1}_t(P_{\min}) - WSDem^{-1}_t(z)
	\end{align}}
where $WSSup$ denotes a supply curve in a wholesale market and $P_{\min}$ shows a minimal ask price available in a market. 

\subsection{{Defining the price classes}}
Having transformed the curves, it is now possible to apply the X-model and carry out price and volume forecasts. Please note that the formulas below are almost identical to those in the original paper. However, the applied transformation allows us to focus only on the supply curve. Therefore, it was possible to omit a great amount of indices in the formulas. Hence, the less sophisticated appearance of the mathematical part of the present paper constitutes another neat simplification of the original X-model. 

To construct the X-model, the first step to be undertaken is to determine price classes on the transformed supply curve. The price classes, as has already been mentioned, are points on the supply curve which correspond to certain prices and to which volume forecasting models will be applied. Then, to proceed further, we first define a grid of prices with positive bid volumes as
\begin{align}
\PP = \{P \in \P|V_t	(P)>0 \}
\end{align}
where $\P$ denotes a grid of all possible possible prices and $V$ stands for volumes on curve $Sup_t^{-1}$. Therefore, the transformed supply curve in our case can be written as
\begin{align}
Sup_t^{-1}(P) = \sum\limits_{\substack{\text{$p \in \PP$} \\ \text{$p \leq P$}} } V_t(p) \text{\hspace{0.2cm} where \hspace{0.2cm}} P \in \PP.
\end{align}
Having determined all possible prices present in the in-sample period, we can proceed with determining the price classes. First of all, we have to compute average volumes over $T$ in-sample observations at prices $\PP$. Hence, it holds that 
\begin{align}
\ov {V}(P) = \frac{1}{T}\sum\limits_{t=1}^{T} V_{t} (P) \text{\hspace{0.2cm} where \hspace{0.2cm}} P \in \PP.
\end{align}
Given the above expression, writing an equation for average curve $\ov{Sup}_t^{-1}$ over the in-sample period yields
\begin{align}
\ov{Sup}^{-1}(P) = \sum\limits_{\substack{\text{$p \in \PP$} \\ \text{$p \leq P$}} } \ov{V}(p) \text{\hspace{0.2cm} where \hspace{0.2cm}} P \in \PP
\end{align}
Finally, we apply an equidistant volume grid with a step of $V_*=500$ mW to curve $\ov{Sup}(iV_*)$. This allows us to define the price classes we are looking for. Hence, it can be said that 
\begin{align}
\C = \{ \ov{Sup} (iV_{*})|i \in \N \}. 
\end{align}

Of course, using an equidistant volume grid may appear too simple for determining the price classes. However, despite its simplicity, the method proves to be relatively powerful. Following Figure \ref{2GraphsInel}, it is clear that the number of price classes is much greater in the flatter segments of the transformed supply curve and is much sparser in the steeper segments. Recall now that the equilibrium price is typically realized in the flatter segments of the curve. Therefore, using the equidistant volume grid allows us to focus only on those sectors of the supply curve in which the price can typically be observed.  Furthermore, please note that there is no mathematical justification for choosing the size of $V_*$. However, the selected value of $V_*$ allows us to to obtain such an amount of price classes that their number is (a) sufficient enough for approximating the supply curve relatively precisely and (b) is not too large and thus not computationally inefficient.

\begin{figure}[h]
	\centering
	\scalebox{0.95}{\input{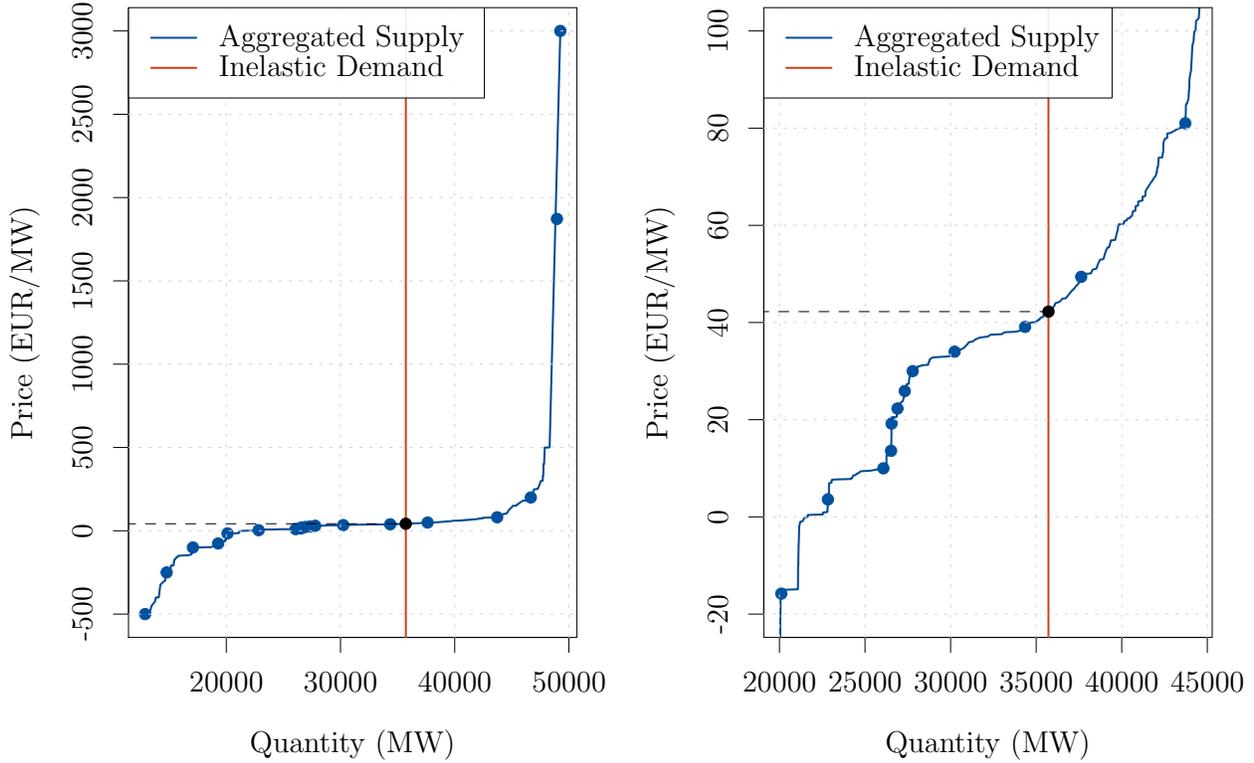}}		
	\caption{A wholesale market equilibrium in the EPEX SPOT SE on 2017-02-01 at 00:00:00 with transformed auction curves and highlighted price classes}
	\label{2GraphsInel}
\end{figure}

Thus, there are $M_S=19$ price classes in our case, as compared to $M_S=16$ in the original paper. The defined price classes $\C$ are represented in Table \ref{TAB1} below. Please note that the volume sizes in each of those price classes $\C$ can be written as 
\begin{align}
X_{S,t}^{(c)}=\sum\limits_{P \in \PP(c)} V_t(P) \text{\hspace{0.2cm} where \hspace{0.2cm}} c \in \C \label{EQVOLUMES}
\end{align}
which means that the volume size in a price class incorporates all volumes present in between this price class and the previous one.

\begin{table}[h!]
	\centering
	{  
		\begin{tabular}{|cccccc|}
			\hline
 			-500.0 &-250.0 &-100.1&  -76.1 & -15.8 &   3.6  \\
 			 10.0  & 13.6  & 19.2  & 22.3 & 25.9  & 30.0  \\
 			  34.0  & 39.1 &  49.4  & 81.0 & 200.0 & 1871.9 \\
 			  3000.0&&&&&\\ 
			\hline 
	\end{tabular}}
	\caption{Price classes $\C$}
	\label{TAB1}
\end{table}

Recall now that the demand curve in our model is perfectly inelastic. Therefore, there are no price classes for the inelastic demand curve. The demand volume is thus given by
\begin{align}
X_{D,t} = Dem_t^{inelastic} \label{EQVOLUMESDEM}
\end{align}
and thus the total amount of variables which we need to forecast equals to $M=M_S+M_D=20$. 

Please note that formulas \ref{EQVOLUMES} and \ref{EQVOLUMESDEM} bear a critical implication for the comparison of the original and the modified X-models. Imagine that we have conducted forecasts for each of $M_S$ price classes and now want to combine the obtained predictions in a single curve. According to formula \ref{EQVOLUMES}, each following point on the forecasted supply curve is an increment over the preceding point. In other words, to construct the predicted supply curve, we have to start with the first price class. Then, to the forecasted volume in the first price class, we one-by-one add the forecasted volumes in the following price classes. Therefore, whenever an outlier occurs in e.g. the 5th price class, this outlier not only affects the 5th price class itself, but also all other price classes afterwards. 

The demand curve in the modified X-model, however, is represented by only one point. Therefore, the modified X-model becomes more robust towards outliers present in the initial auction curves data. This holds since the cumulative effect of outliers does not affect the demand forecast in the modified X-model. Hence, accuracy of the modified X-model is higher compared to that of the original X-model.

\subsection{Time series model} 
Please note that the applied time series model is very similar to the one in the original paper by \cite{ziel2016electricity} and is thus similar to \cite{weron2008forecasting} or \cite{ziel2016forecasting}. The model is a simple ARX-type process with 4 external regressors. The external regressors are the wholesale market price and forecasts for: electricity generation, wind and solar power supply. Please note that in our case the equilibrium volume coincides with the value of the inelastic demand function. Therefore, to account for the equilibrium volume separately, we consider the difference between the equilibrium volume in the setting of the transformed curves and the initial wholesale equilibrium volume, i.e. $X_{d,h}^{volume}=X_{d,h}^{Dem^{inelastic}_{d,h}}-X_{d,h}^{volume, WS}$. The time series for the modified X-model with an inelastic demand curve is then
	\begin{align}
X_{d,h} = \left( \left(X_{S,d,h}^{(c)}\right)_{c \in \C}, X_{D,d,h}, X_{d,h}^{price}, X_{d,h}^{volume}, X_{d+1,h}^{generation}, X_{d+1,h}^{wind}, X_{d+1,h}^{solar} \right)
\end{align}
which in this case includes $M+5=25$ variables. However, the forecast is conducted only for the first $M_S+M_D=20$ parameters since the remaining ones are only auxiliary. 

To capture the seasonal structure, a weekday dummy is introduced with the following formula
\begin{align}
W_k(d)\begin{cases}
1, \hspace{0.5cm}\WW(d)<k\\
0, \hspace{0.5cm} \WW(d)\geq k 
\end{cases}
\end{align}
where $\WW(d)$ is a function which yields a number corresponding to the day of the week $d$ and $k$ is a day index with e.g. $k=1$ for Monday. 

Since we estimate the time series model by a BIC-based lasso (for more see \cite{tibshirani1996regression} and \cite{schwarz1978estimating}), the underlying data should be standardized. Therefore, we have to subtract means from the original process, i.e. $\bY_{d,h} = \bX_{m,d,h}-\bmu_{h}$ where $\bmu_h=\E(\bX_{d,h})$. Therefore, the model under consideration can be written as follows
\begin{align}
Y_{m,d,h} = \sum\limits_{l=1}^M \sum\limits_{j=1}^{24} \sum\limits_{k\in \II_{m,h}(l,j)} \phi_{m,h,l,j,k} Y_{l,d-k,j} + \sum\limits_{k=2}^7 \varphi_{m,h,k} W_k (d) + \varepsilon_{m,d,h}
\end{align}
where $\phi_{m,h,l,j,k}$, $\varphi_{m,h,k}$ and $\II_{m,h}(l,j)$ are sets of lags and $\varepsilon_{m,d,h}$ is an error term. As in the original paper, the latter term is supposed to be i.i.d. with constant variance $\sigma^2_{m,h}$. Please note that $\II_{m,h}(l,j)$ is defined as 	
\begin{align}
\II_{m,h}(l,j) = \begin{cases}
\{1,2,3,...,36\}, \hspace{0.2cm} m=l \text{\hspace{0.1cm} and \hspace{0.1cm} } h=j \\
\{1,2,3,...,8\}, \hspace{0.2cm} (m=l \text{\hspace{0.1cm} and \hspace{0.1cm} } h\neq j) \text{\hspace{0.1cm} or \hspace{0.1cm} } (m \neq l \text{\hspace{0.1cm} and \hspace{0.1cm} } h=j) \\
\{1\},\hspace{0.2cm} m \neq l \text{\hspace{0.1cm} and \hspace{0.1cm} } h\neq j 
\end{cases}
\end{align}
where the choice of lags and the corresponding motivation is elaborated at length in the original paper by \cite{ziel2016electricity}. 

Then, to estimate the $\beta$-coefficients, we use R-package \texttt{glmnet} (for more see e.g. \cite{friedman2010regularization}). The multivariate ordinary least squares estimator for our model can thus be defined as 
\begin{align}
\wtilde{Y}_{m,d,h} = \wtilde{\X}_{m,d,h} \wtilde{\bbeta}_{m,h} + \wtilde{\varepsilon}_{m,d,h}
\end{align}
where $\X_{m,d,h} = (\X_{m,d,h,1},...,\X_{m,d,h,p_{m,h}})'$ is a $p_{m,h}$-dimensional vector of regressors, $\bbeta_{m,h}$ is a corresponding vector of coefficients, and tilde denotes a standardized version of a variable with its variance being scaled to one. Standardization is necessary for the lasso-estimator to function correctly. 	
The corresponding mathematical representation of the scaled and estimated $\what{\wtilde{\beta}}$-coefficients can be written as follows
\begin{align*}
\what{\wtilde{\bbeta}}_{m,h} = \argmin_{\bbeta \in \R^{p_{m,h}}} \sum\limits_{d=1}^n \left(\wtilde{Y}_{m,d,h} - \bbeta\wtilde{\X}_{m,d,h} \right)^2 + \lambda_{m,h} \sum\limits_{j=1}^{p_{m,h}} |\bbeta_j|
\end{align*}
where $\lambda_{m,h}$ denotes a penalization parameter. Moreover, please note that the non-standardized versions of the coefficients can be obtained easily by rescaling.

The volume forecast for the next day is thus given by
\begin{align}
\what{Y}_{m,n+1,h} = \sum\limits_{l=1}^M \sum\limits_{j=1}^{24} \sum\limits_{k\in \II_{m,h}(l,j)} \what{\phi}_{m,h,l,j,k} Y_{l,n+1-k,j} + \sum\limits_{k=2}^7 \what{\varphi}_{m,h,k} W_k (n+1).  
\end{align}
Then, we need to add sample means to the obtained values of $\what{Y}_{1,n+1,h},...,\what{Y}_{M,n+1,h}$ to compute the final day-ahead volume forecast $\what{X}_{1,n+1,h},...,\what{X}_{M,n+1,h}$. Please note, however, that to calculate a precise forecast for the next day simply adding mean values to the above defined process is not sufficient. We thus follow the procedure used in the original paper and run a residual-based bootstrap simulation with $B=10000$ bootstrap samples. Hence, we sample from the residual vector $\what{\beps}_{d,h} = (\varepsilon_{1,d,h},...,\varepsilon_{M,d,h})'$ only over the days $d$. We then use the mean of the simulated results to finalize the computation of our point forecasts. 

\subsection{Supply curve reconstruction}
The model described in the previous section is then applied to each of the price classes. As a result, we have day-ahead forecasts for $M_S=19$ points which lie on the forecasted supply curve and a forecast for the inelastic demand. Therefore, what remains to be done is to connect the forecasted $M_S$ price classes with each other, i.e. draw a curve out of the prognosticated points. We, however, want to retain the structure of the transformed supply curves and thus want to replicate this structure as precisely as possible. Hence, we do not simply draw a line over the predicted points, but instead use a more sophisticated technique. This technique was called curve reconstruction in the original paper. We thus rely on this technique without further modifying it. 

\begin{figure}
	\centering
	\vspace{-0.7cm}
	\scalebox{0.9}{\input{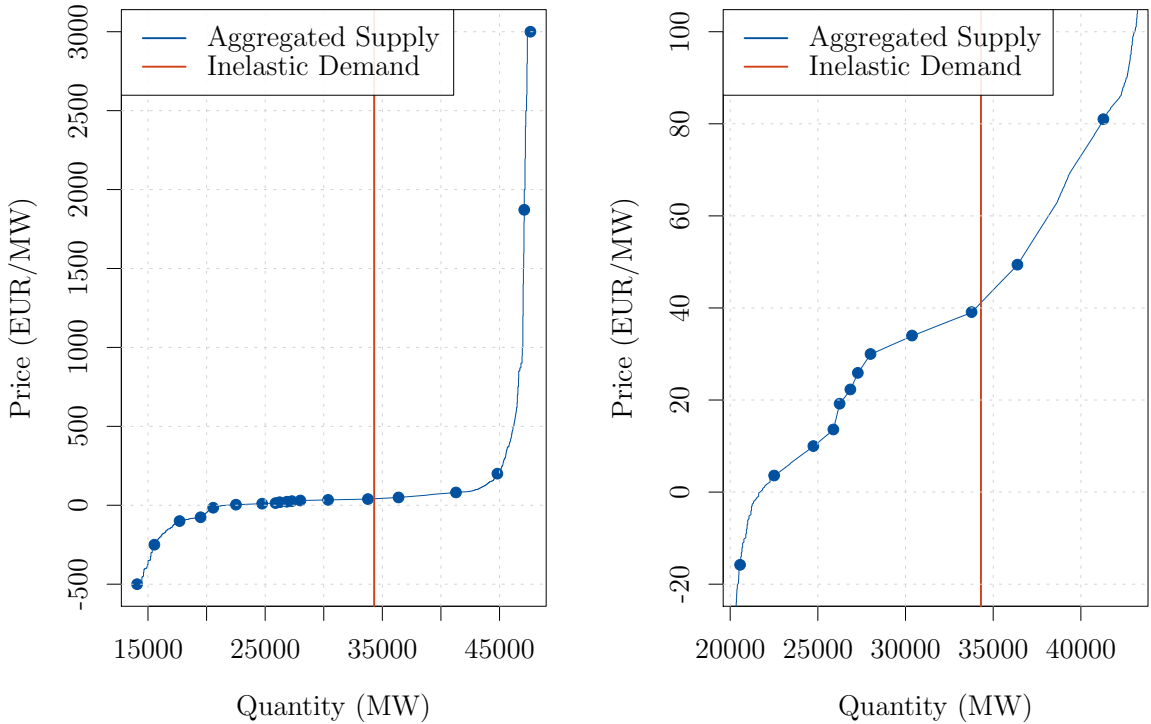}}
	\vspace{-0.2cm}
	\caption{{Supply curve reconstruction on 2017-02-01 at 00-00-00}}
	\label{3Graphs}
\end{figure}

To proceed further, we consider the following formula
	\begin{align}
\check{V}_{d,h}(P) = \frac{R(P)\ov{V}(P)}{\sum_{Q \in \PP(c)}R(Q)\ov{V}(Q)} X_{d,h}^{(c)} \label{CURVEREC}
\end{align}
where $R(P)=1$ if a price occurs at least two times a day and $R(P)=0$ otherwise. Equation \ref{CURVEREC} thus allows us to neglect prices which are not important and hence models the actual composition of the supply curve more accurately. Please note that reconstructing the demand curve is not necessary since this curve is perfectly inelastic. Figure \ref{3Graphs} provides a graphical representation of the prognosticated curves after the curve reconstruction was carried out. 

\section{The obtained results}
To test the model, a rolling window study was conducted. The size of the window was equal to one day, whereas the out-of-sample period was equal to the year 2017. The comparison between the modified and the original X-models with the naive benchmark is provided in the Table \ref{TAB2} below. Please note that the definitions of the MAE- and RMSE-values are analogous to those in the original paper or in e.g. \cite{uniejewski2017variance}. 

	\begin{table}[h!]
	\centering
	\begin{tabular}{|c|cc|c|}
		\hline
		&MAE&RMSE& Average execution time (min)\\
		\hline
		Naive&9.97 & 11.90 & - \\
		X-model orig. &6.21 &7.54 & 4.34 \\
		X-model inel. &5.12 & 6.45 & 1.40\\
		\hline 
	\end{tabular}
	\caption{Comparison of the naive benchmark vs. the original X-model (\cite{ziel2016electricity}) vs. the modified X-model with an inelastic demand curve}
	\label{TAB2}
\end{table}

Besides lower MAE- and RMSE- values, the conducted DM-test has also proven superiority of the modified X-model with the corresponding $p$-value being equal to $2\times10^{-9}$. Therefore, following Table \ref{TAB2} and the previous discussion, the modified X-model outperforms the original one in two major aspects. 

The first aspect is the execution speed. The modified X-model requires on average 1.4 minutes to deliver the results, whereas the original model needs on average 4.34 minutes. Please note that the execution speed may vary depending on the specification of the lasso model and its parameters. Yet, the obtained results demonstrate explicitly that the modified version of the X-model is significantly faster. Naturally, the improvement occurs because the amount of variables is almost twice smaller in the modified version of the X-model. 

The second aspect is the quality of results. As has been mentioned earlier, the modified X-model is more robust towards outliers present in the initial auction curves data because the cumulative effect of outliers is absent in the modified version of the X-model. Naturally, this means that the demand curve is approximated more accurately in the modified version of the X-model. In turn, this leads to a significant improvement of the accuracy.  

\begin{figure}[h!]
	\centering
	\scalebox{0.95}{\input{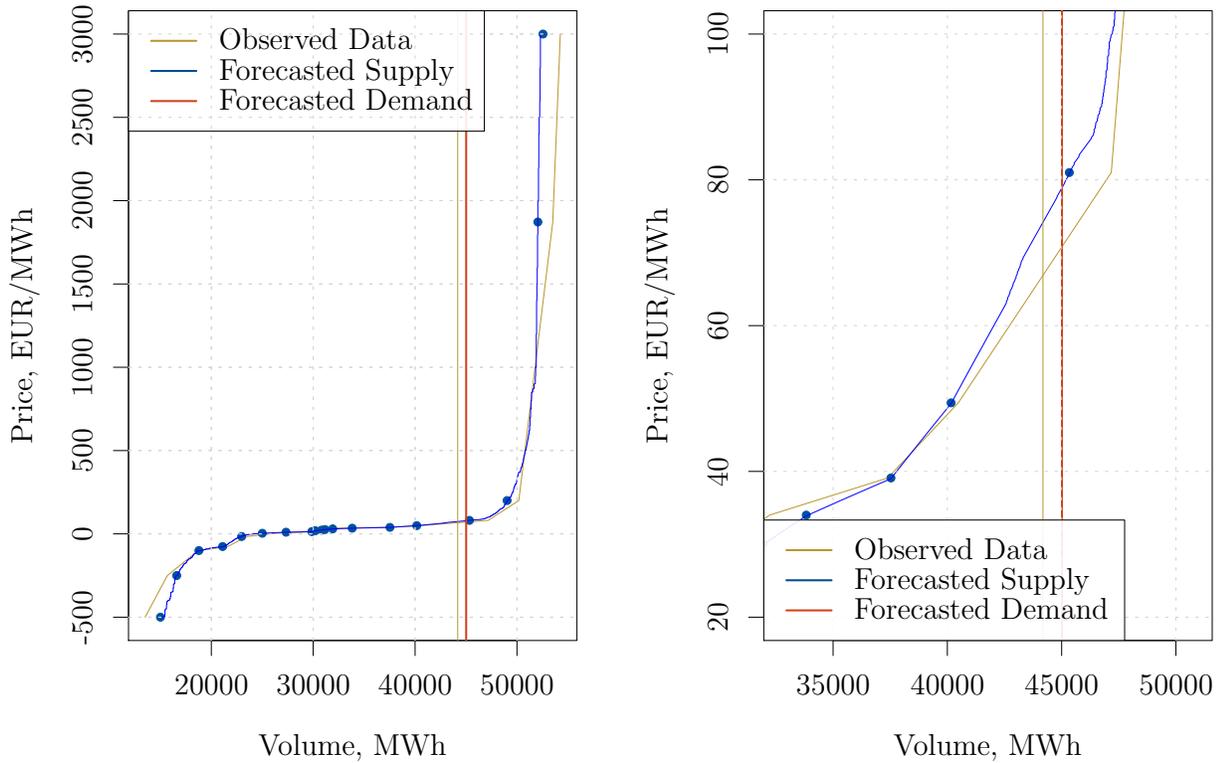}}
	\caption{{Market equilibrium forecast on 2017-02-01 at 10-00-00}}
	\label{Results}
\end{figure}

Figure \ref{Results} shows the forecast for the supply and demand curves (depicted in blue and red, respectively) delivered by the modified X-model against the true data (depicted in yellow). 

Moreover, an example of the equilibrium price and volume forecasts can be seen in Figure \ref{ForecastedPricesVolumes} below. As can be seen explicitly, the modified X-model proves it suitability for both volume and price forecasting. Moreover, as the above discussion demonstrates, the modified X-model is superior to the original one in both quality and speed dimensions. 
\begin{figure}[h!]
	\centering
	\scalebox{0.9}{\input{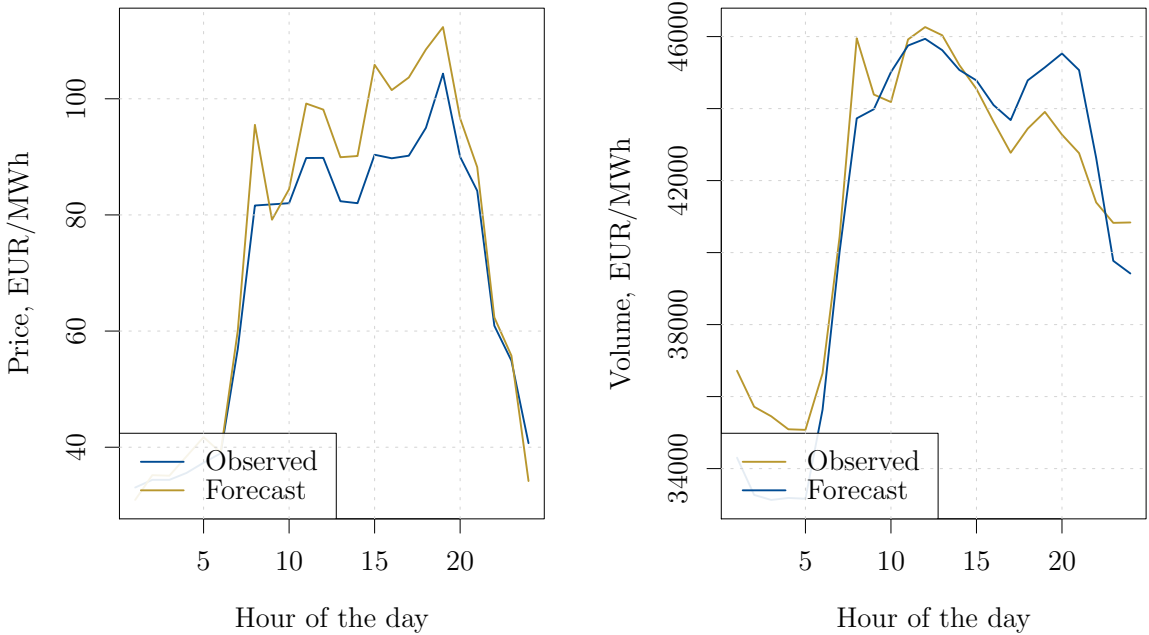}}
	\caption{{Prices and volumes forecast on 2018-01-31}}
	\label{ForecastedPricesVolumes}
\end{figure}

\section{Conclusion}
The core idea of the present paper was to provide an improvement to the famous X-model derived by \cite{ziel2016electricity}. Since the X-model has not been widely studied as yet, the present paper is presumably the first one which develops the X-model further. 

The key component of the improvement came from the transformation of the auction curves into their analogues with an inelastic demand curve. The fundamentals behind the transformation were taken from \cite{coulon2014hourly}. We showed that using this method prior to applying the X-model leads to a significant improvement of the final results. 

More specifically, the modified X-model was shown to work faster. The boost in the execution speed came from the fact that the demand curve after the transformation was represented by only one point instead of several price classes. Therefore, predicting the demand curve in the modified X-model is much less computationally expensive. 

Moreover, the modified X-model was shown to be more robust towards outliers present in the initial auction curves data. Due to the specifics of the model, these outliers may influence the compositions of the forecasted curves significantly. This influence, in turn, may deteriorate the quality of price and volume forecasts. Since it is much simpler to predict the demand curve in the modified version of the X-model, outliers exert a weaker influence upon the model's precision. Therefore, the modified X-model yields more accurate forecasts.

\newpage

\newpage 
\bibliographystyle{apalike}
\bibliography{lib}

\end{document}